\PassOptionsToPackage{svgnames,dvipsnames,table}{xcolor}
\documentclass[a4paper,fleqn]{cas-dc}

\usepackage[numbers]{natbib}
\usepackage{todonotes}
\usepackage{csquotes}
\usepackage[width=\textwidth]{caption}

\usepackage[dvipsnames]{xcolor}

\definecolor{myBlue}{rgb}{0, 0, 255}
\definecolor{myRed2}{rgb}{127, 19, 1}
\definecolor{myGreen}{rgb}{0, 97, 0}

\def\tsc#1{\csdef{#1}{\textsc{\lowercase{#1}}\xspace}}
\tsc{WGM}
\tsc{QE}
\tsc{EP}
\tsc{PMS}
\tsc{BEC}
\tsc{DE}
\RequirePackage[normalem]{ulem} 
\RequirePackage{color}\definecolor{RED}{rgb}{1,0,0}\definecolor{BLUE}{rgb}{0,0,1} 

\begin{document}

\let\WriteBookmarks\relax
\def\floatpagepagefraction{1}
\def\textpagefraction{.001}
\shorttitle{Crowdsourced Behavior-Driven Development}
\shortauthors{Emad Aghayi et~al.}

\title [mode = title]{Crowdsourced Behavior-Driven Development}                      
\tnotemark[1,2]

\tnotetext[1]{This research
   project was supported in part by the National Science Foundation under grants CCF-1414197 and CCF-1845508.}

\tnotetext[2]{Crowd Microservices: Crowdsourced Behavior-Driven Development}


\author[1]{Emad Aghayi}[type=editor, auid=000,bioid=1,  orcid=0000-0003-0607-4257]
\cormark[1]
\ead{eaghayi@gmu.edu}
\ead[url]{www.mason.gmu.edu/~eaghayi}


\address[1]{Department of Computer Science, George Mason University, 4400 University Drive, Fairfax, VA 22030}

\author[1]{Thomas D. LaToza}
\cormark[2]
\ead{tlatoza@gmu.edu}
\ead[URL]{www.cs.gmu.edu/~tlatoza}

\author[1]{Paurav Surendra}
\ead{psurendr@gmu.edu}

\author%
[2]{Seyedmeysam Abolghasemi}
\ead{sabolgha@cs.odu.edu}
\address[2]{Old Dominion University, 1112 Monarch Hall, Norfolk, VA 23529}

\cortext[cor1]{Corresponding author}
\cortext[cor2]{Principal corresponding author}


\begin{abstract}
Key to the effectiveness of crowdsourcing approaches for software engineering is workflow design, describing how complex work is organized into small, relatively independent microtasks. In this paper, we introduce a Behavior-Driven Development (BDD) workflow for accomplishing programming work through self-contained microtasks, implemented as a preconfigured environment called \textit{Crowd Microservices}. In our approach, a client, acting on behalf of a software team, describes a microservice as a set of endpoints with paths, requests, and responses. A crowd then implements the endpoints, identifying individual endpoint behaviors which they test, implement, and debug, creating new functions and interacting with persistence APIs as needed. To evaluate our approach, we conducted a feasibility study in which a small crowd worked to implement a small \textit{ToDo} microservice. The crowd created an implementation with only four defects, completing 350 microtasks and implementing 13 functions. We discuss the implications of these findings for incorporating crowdsourced programming contributions into traditional software projects.  
\end{abstract}

\begin{graphicalabstract}
\includegraphics[width=\textwidth,keepaspectratio, clip]{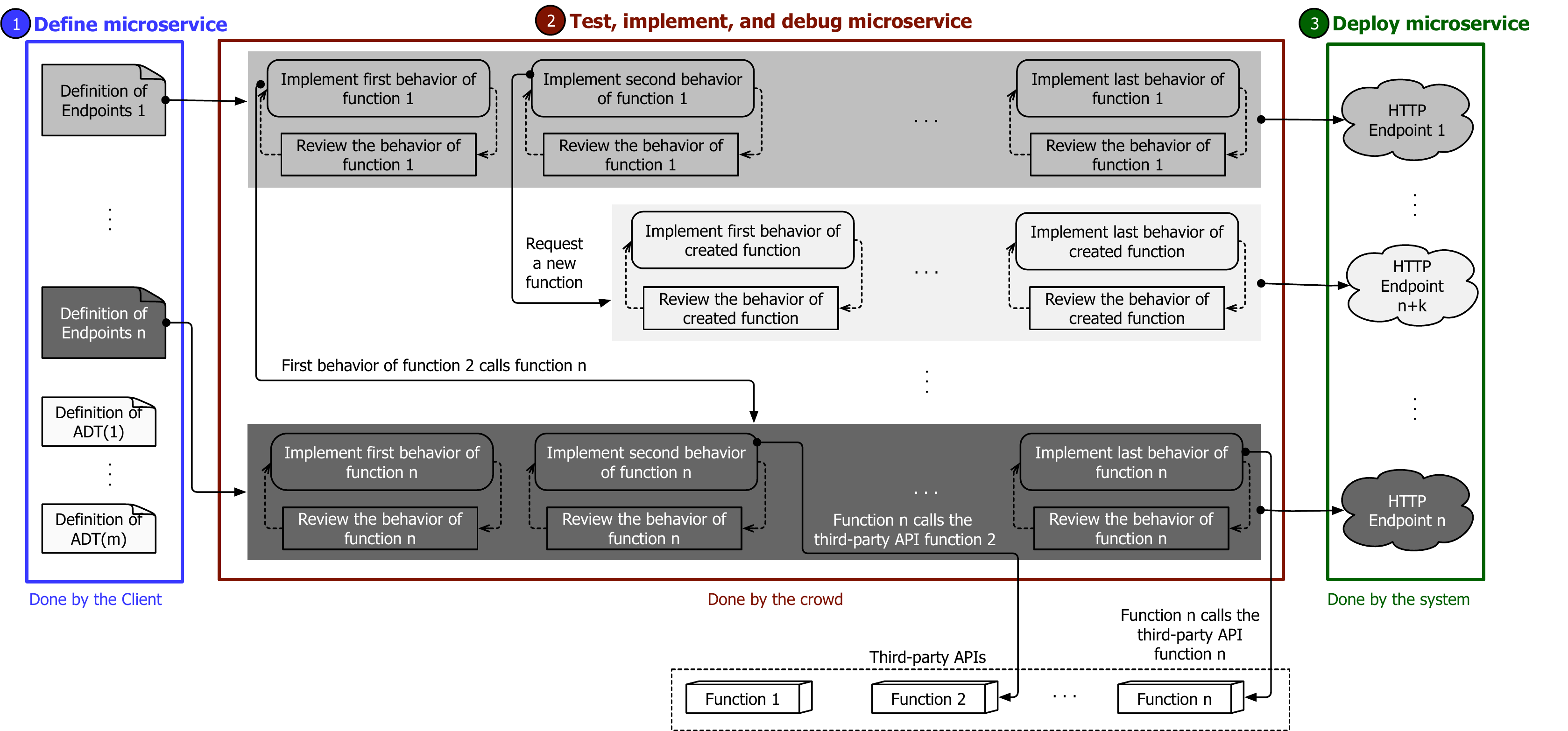}

\end{graphicalabstract}

\begin{highlights}
\item What if the microtasking model could be applied to software development?
\item A novel BDD approach for enabling immediate feedback during programming microtasks.
\item Crowd Microservices: first IDE for implementing microservices through microtasks. 
\item Crowd Microservices demo: \url{ https://www.youtube.com/watch?v=qQeYOsRaxHc&t=17s}
\item An exploratory study to evaluate Crowd Microservices and its workflow
\end{highlights}


\maketitle

\section{Introduction}

Crowdsourcing software engineering offers developers new ways to contribute to software projects, enabling workers outside a traditional software development team to take part in building software~\cite{latoza2016crowdsourcing}. 
One form of crowdsourcing is microtask crowdsourcing, in which a workflow is used to decompose tasks into a sequence of microtasks. By reducing the necessary task context for newcomers to learn,  microtask programming reduces the cost of onboarding and enables developers to contribute more quickly. Small contributions also open the possibility of increasing parallelism in software development. As many hands make light work, decomposing software development tasks into microtasks might enable some of this work to be completed in less time by parallelizing work across many developers.

A variety of systems have explored the promise of applying microtask crowdsourcing to programming \cite{codeon2017,Collabode:2011,Collabode:2012,lasecki2015apparition,latoza2015borrowing,latoza2018microtask,LaToza2015Scaffolding,latoza2014microtask,latoza2013crowd,crowddesign,schiller2012reducing,MAO2017survey}.
For example, in Apparition~\cite{lasecki2015apparition}, a client developer narrates a description for a user interface in natural language, and crowd workers translate this description into user interface elements, visual styles, and behavior. In CodeOn~\cite{codeon2017}, a client developer narrates a request for help from their IDE, and crowd workers use this request and relevant context to author answers and code.

A long-term vision of microtask programming is to enable software to be built entirely through microtasks~\cite{latoza2013crowd}. This differs from current approaches in scale: rather than facilitating individual tasks to be manually requested and managed by a requester, mechanisms must be found for the crowd to coordinate and work more effectively together. In microtasking, required coordination is described through a workflow, describing the microtasks which developers will complete, the handoffs that occur between microtasks, and the resulting dependencies between microtasks
~\cite{kittur2013future}.
A key challenge in applying microtask crowdsourcing approaches to the domain of software engineering is the decontextualized nature of microtask work. Developers work without awareness of the complete program, reducing the necessary context which must be learned to contribute but increasing the potential for work going off track. Key to the success of microtasking approaches is to ensure effective mechanisms for coordination and aggregation exist, enabling workers to obtain feedback as quickly as possible to ensure their contributions are beneficial and coordinate contributions so that they do not conflict. When this does not occur, problems may ensue. Developers may write code which references fields and functions that do not exist, making poor implementation choices, or making implementation choices which conflict with other choices~\cite{latoza2014microtask,latoza2018microtask}. As a result, time and effort are wasted, as further work is required to fix these issues, and the potential for undetected defects increases. 

In this paper, we introduce a new approach for feedback based on behavior-driven development. In contrast to existing approaches which rely on either a client or manager~(e.g., \cite{lasecki2015apparition,Collabode:2012}) or later crowd contributions~(e.g., \cite{latoza2014microtask,latoza2018microtask}), we enable developers to receive initial feedback within the microtask itself. We accomplish this by re-envisioning the scope of the microtask. In our workflow, developers receive feedback from three different sources while they are programming: through syntax errors, through running unit tests, and through the ability to debug their code.
Adapting the idea of behavior-driven development to crowdsourcing work ~\cite{north2006introducing,beck2003test}, each microtask encompasses the work to test and implement a \textit{behavior}, a specific identifiable use case of a function. Developers work on a behavior end-to-end, identifying it from a high-level description of a function, writing a test to exercise it, implementing it in code, and debugging any issues that emerge. Through the use of stubs, developers can work on an individual function in isolation from the rest of the program being constructed while still receiving information about how their code executes. As developers join a new project, they complete a tutorial and can immediately begin making small contributions to the project.

We apply our behavior-driven approach to microtask programming to implementing a microservice. Large and complex web applications are often built as an interconnected network of smaller single-function microservices.
Microservices offer a well-defined interface between a client (e.g., a team consuming a future microservice) and a crowd of developers, defined by the behavior of a set of HTTP endpoints handled by the microservice. 
Microservices offer a natural decomposition boundary in large web applications, offering a mechanism for a traditional software project to quickly crowdsource a module within a larger project. 
Rather than require an extended onboarding process for new developers to become familiar with a large software project, our approach enables a team to simply describe the desired functionality and it to be constructed separately by a crowd.

We instantiated our approach in a novel cloud IDE, \textit{Crowd Microservices}\footnote{https://youtu.be/mIn2EOqsDYw}.~\textit{Crowd Microservices} includes an editor for clients to describe requirements for the system as a set of endpoints, a web-based programming environment in which crowd workers can identify, test, implement, and debug behaviors, and an infrastructure for automatically generating and managing microtasks. Persistence is exposed through an external API, which supports a sandboxed environment for development. After completion, the microservice may be automatically deployed to a hosting site. 

To evaluate our approach, we conducted a user study in which 9 crowd workers implemented a simple~\textit{ToDo} microservice. Our results offer initial evidence for the feasibility of the approach. Participants submitted their first microtask 24 minutes after beginning, successfully submitted 350 microtasks, implemented 13 functions and 36 tests, completed microtasks in a median time under 5 minutes, and correctly implemented 27 of 34 behaviors.

In this paper, we contribute
\begin{enumerate}
    \item a novel behavior-driven microtask programming which offers immediate feedback from syntax errors, unit tests, and debugging.
    \item \textit{Crowd Microservices}, the first programming preconfigured environment for implementing microservices through microtasks
    \item initial evidence that behavior-driven microtasks can be quickly completed by developers and used to successfully implement a microservice.
    \item evidence that the approach has low onboarding time and a high potential for parallelism.
 \end{enumerate}
 In the rest of the paper, we review related work, present our approach, illustrate the approach with an example, and report on a user study evaluation. We conclude with a discussion of limitations and threats to vlaidity as well as opportunities and future directions.\par
 

\begin{table*}
\centering

\caption{Applying the dimensions of crowdsourcing to concrete collaboration examples.}
\label{tab:relatedWork}
\resizebox{\linewidth}{!}{%
\rowcolors{1}{gray!40}{}
\begin{tabular}{lllllllllll}
\multicolumn{1}{l}{} &
\multicolumn{1}{l}{\begin{tabular}[c]{@{}c@{}}\textbf{De-} \\ \textbf{composition} \end{tabular}}& 
\multicolumn{1}{l}{\begin{tabular}[c]{@{}c@{}}\textbf{Task Inter-} \\ \textbf{dependence} \end{tabular}} &
\textbf{Task context} & \textbf{Task length} & \textbf{Activities} & \textbf{Onboarding} & 
\multicolumn{1}{l}{\begin{tabular}[c]{@{}c@{}}\textbf{Quality} \\ \textbf{control} \end{tabular}}&
\multicolumn{1}{l}{\begin{tabular}[c]{@{}c@{}}\textbf{Locus of} \\ \textbf{control} \end{tabular}} &

\multicolumn{1}{l}{\begin{tabular}[c]{@{}c@{}}\textbf{Preconf-} \\ \textbf{igured IDE} \end{tabular}}  \\

\textbf{Crowd Microservices} & Automatic & Low & \multicolumn{1}{l}{\begin{tabular}[c]{@{}c@{}} A description, fuction\\and unit tests\end{tabular}}  & <= 5 min  & Crowd Develop/debug/test  & <= 15 min &
\multicolumn{1}{l}{\begin{tabular}[c]{@{}c@{}}unit tests \& \\  reviews by crowd \end{tabular}}  & Client & Yes \\
\textbf{CrowdCode~\cite{latoza2018microtask}} & Automatic & Low & A function  & <= 5 min  & Crowd Develop/debug/test  & <= 15 min &
\multicolumn{1}{l}{\begin{tabular}[c]{@{}c@{}}unit tests \& \\  reviews by crowd \end{tabular}}  & Client & Yes \\

\textbf{Apparition~\cite{lasecki2015apparition}} & Manual & Low & Whole design& <= 1 minute&  Real-time UI Design & - &None & Designer &Yes \\

\textbf{CrowdDesign~\cite{crowddesign}} & Manual & Low &\multicolumn{1}{l}{\begin{tabular}[c]{@{}c@{}} An statement\\to a module\end{tabular}} &  <=15 min &Implement UI element&-&\multicolumn{1}{l}{\begin{tabular}[c]{@{}c@{}}reviews by \\ manager \end{tabular}}&manager&Yes  \\

\textbf{Crowdforge~\cite{kittur2011crowdforge}}&Manual &Low&a partition&Low - high& crowdsourcing complex tasks&-&
\multicolumn{1}{l}{\begin{tabular}[c]{@{}c@{}}reviews by \\ manager \end{tabular}} &
Manager&Yes \\

\textbf{CodeOn~\cite{codeon2017}} & Manual &Low & 
\multicolumn{1}{l}{\begin{tabular}[c]{@{}c@{}}An statement\\to whole codebase\end{tabular}} &
<= 11 minutes & Help Seeking in  Development&-  &Node & Requester &Yes \\

\textbf{Collabode~\cite{Collabode:2012}} & Manual & High & Whole codebase & Min - days & \multicolumn{1}{l}{\begin{tabular}[c]{@{}c@{}}Synchronous collaborative \\ coding\end{tabular}} &-&None&Collaborators&Yes \\

\textbf{CodePilot~\cite{CodePilot:2017}}& Manual&High&Whole codebase& Min - hours & 
\multicolumn{1}{l}{\begin{tabular}[c]{@{}c@{}}Synchronous collaborative \\ Coding for novices\end{tabular}}  &- &Unit testing& Collaborators & Yes \\

\textbf{Open-source} & Manual &Medium-High&Whole codebase&Hours - days& Crowd Develop/debug/test &Hours - days &\multicolumn{1}{l}{\begin{tabular}[c]{@{}c@{}}Unit tests \& \\  Reviews by crowd \end{tabular}}  &
\multicolumn{1}{l}{\begin{tabular}[c]{@{}c@{}} Senior \\ contributors \end{tabular}} & No \\

\textbf{TopCoder}&Manual &Low& \multicolumn{1}{l}{\begin{tabular}[c]{@{}c@{}} A function\\to a module\end{tabular}}&Minutes - hours&  Design/Development&Minutes - hours& Manager & \multicolumn{1}{l}{\begin{tabular}[c]{@{}c@{}}Reviews by \\ manager \end{tabular}} & No \\

\end{tabular}
}
\end{table*}

\section{Related work}
Our work builds on a broad body of work in crowdsourcing, particularly prior approaches to crowdsourcing software development. This work has investigated several key aspects of microtask workflow design, which we focus on below: decomposition and context, parallelism and conflicts, enabling fast onboarding, and achieving quality. A summary of these approaches is presented in Table~\ref{tab:relatedWork}. 

\subsection{Decomposition, context, and scope}
A key challenge in microtasking in all domains is the manner in which work is decomposed into the tasks or microtasks which are completed by crowd workers. The choice of decomposition determines the workflow of the approach, encompassing the individual steps, the context and information offered in each step, and the types of contributions which can be made~\cite{retelny2017noworkflow,kittur2011crowdforge,bernstein2010soylent,hoseini2018multidimensional,kittur2013future,jiang2014efficient}.\par

Within approaches for crowdsourcing software engineering work, several points in the design space have been explored~\cite{MAO2017survey}. Depending on the choice of microtask boundaries, contributions may be easier or harder, may vary in quality, and may impose differing levels of overhead. Approaches to crowdsourcing programming explore novel workflows designed specifically for accomplishing specific software development tasks.  \par

Techniques for decomposing programming work into fine-grained microtask may be either manual or automatic (Table~\ref{tab:relatedWork}). Manual approaches rely on a developer or client to author each microtask. For example, in CodeOn~\cite{codeon2017}, a developer working in a project requests small microtasks for others to complete.  In Apparition~\cite{lasecki2015apparition}, a similar workflow is used to build user interface prototypes. Requestors describe desired user interface elements and their user interaction through natural language todo items. Workers then view the output of the complete program, select a microtask from the todolist, and implement the requested functionality for a UI element. While working on the microtask, workers interact with an individual element, but otherwise have a global view of the entire codebase. In Crowd Design~\cite{crowddesign}, work for building a web application is broken down by component. Crowd members work individually in an isolated environment on each component and complete design and testing tasks.
Manual decomposition of work limits the scalability of a crowdsourcing system. As microtasks are generated manually by a single individual with a global view of the project, scalability is limited.

Other systems automatically generate microtasks from work finished previously by the crowd, reducing the work imposed on the client to create and manage microtasks. In CrowdCode, programming work is done through a series of specialized microtasks in which participants write test cases, implement tests, write code, look for existing functions to reuse, and debug \cite{latoza2014microtask,latoza2018microtask}. Other work has automatically generated microtasks through puzzle games, formulating complex tasks such as testing and verification as puzzles which can be completed with little or no programming knowledge \cite{chen2012puzzle,schiller2012reducing,lerner2015polymorphic,Bounov:2018:ILI:3173574.3173805}. \par

Task interdependence relates to the degree to which a task requires organizational units to affect the activities and work outcomes of other units~\cite{goldman1977intergroup,andres2002contingency}. Increasing task interdependence forces team members to integrate their contributions with others. When task interdependency is low, the contributions of each unit or team member is additive. There exists a relation between the degree of task interdependency and the productivity of units or team members~\cite{straus1994does}. Higher degrees of task interdependencies require more information exchange, more task clarification in task assignment, creating more shared mental models, and more effort to integrate tasks~\cite{straus1994does}. Productivity decreases because of time spent establishing shared mental models and resolving conflicts among contributors. Studies have shown productivity dramatically declines when a team member needs the output of other units or team members as an input to their task~\cite{andres2002contingency}. In designing microtask workflows, it can thus be beneficial to decrease the degree of the interdependency of software development tasks by creating tasks with scopes that developers can work on with less interdependency. In our workflow, microtasks are largely self-contained and independent of the other microtasks, which provided a low level of dependency. Moreover, implementing tests and implementing behaviors are completed by the same crowd worker, which helps to avoid spending time to resolve conflicts.

The context of a task in crowdsourcing software development systems may vary from several statements in a function to the whole codebase (Table~\ref{tab:relatedWork}). For instance, in Apparition, each contributor can see the entire codebase to complete a task. In \textit{Crowd Microservices}, the context is a function description, a function, unit tests of individual behavior.


\subsection{Parallelism and conflicts}
By decomposing large tasks into smaller tasks, work can be assigned to several workers and completed more quickly in parallel. For easily parallelizable software engineering tasks, like writing a test, this paradigm has achieved widespread adoption in commercial platforms. Crowdsourced testing platforms such as UserTesting \footnote{https://www.usertesting.com}, TryMyUI \footnote{https://www.trymyui.com}, and uTest\footnote{https://www.utest.com} enable software projects to crowdsource functional and usability testing work by utilizing the crowd of tens or hundreds of thousands of contributors on these platforms.\par 

Microtasking approaches for programming envision reducing the necessary time to complete programming tasks through parallelism. A key challenge occurs with conflicts, where two overlapping changes to an artifact are submitted at the same time. For example, in traditional software development, each contributor may edit the same artifacts at the same time, resulting in a merge conflict when conflicting changes are committed. This is an example of an optimistic locking discipline,  where any artifact may be edited by anyone at any time. Due to the increased parallelism assumed and greater potential for conflicts,  microtasking approaches often apply a pessimistic locking discipline, where microtasks are scoped to an individual artifact and further work on these artifacts is locked while they are in progress. For example, in Apparition \cite{lasecki2015apparition} workers acquire write-locks to avoid conflicts. 
Similarly, in CrowdCode \cite{latoza2018microtask} contributions are made on an individual function or test, which is locked while a microtask is in progress. However, conflicts may still occur when decisions made in separate microtasks must be coordinated and are not \cite{Zanatta:2018:Topcoder,LaToza2015Scaffolding}. For example, conflicts may occur when microtasks are separately required to translate function descriptions into an implementation and tests and differing interpretations of a function description are made \cite{latoza2018microtask}. 
In this paper, we explore an approach for limiting conflicts by decomposing work around behaviors, removing the potential for conflicts caused by tests and code written by different developers. 
 \par




\subsection{Achieving quality}
Crowdsourcing approaches to programming have explored a variety of approaches for ensuring the quality of the resulting software artifacts. The quality is ultimately determined by the quality of individual contributions. There are many causes of low-quality contributions, including workers who do not have sufficient knowledge, who put forth little effort, or who are malicious \cite{Kim:2017:MNC:2998181.2998196}. A study of the TopCoder\footnote{https://www.topcoder.com/} crowdsourcing platform revealed six factors related to project quality, including the 
average quality score on the platform, the number of contemporary projects, the length of documents, the number of registered developers, the maximum rating of submitted developers, and the design score\cite{li2013analysis}. In TopCoder, senior contributors assist in managing the process of creating and administering each task and ensuring quality work is done~\cite{Stol2014:TwoCompany}. 
Studies of software crowdsourcing companies identified 10 methods used for addressing quality, including ranking and ratings, reporting spam, reporting unfair treatment, task pre-approval, and skill filtering~\cite{saengkhattiya2012quality}. \par


In microtask systems, crowd members are often assumed to be minimally invested in the platform or community.  Crowd programming systems have addressed this problem by assigning the responsibility of feedback and management to the client or the developers requesting the work \cite{Collabode:2012,codeon2017,lasecki2015apparition,lasecki2018apparition2}. Systems where the requestor is less directly involved in work and microtasks are automatically generated may have crowd members review and offer feedback after contributions are made~\cite{latoza2018microtask}. However, this approach is limited, as contributors who do not receive the traditional feedback offered in programming environments, such as syntax errors, missing references, and unit test failures, may submit work which contains these issues, which other contributors must then address later at higher cost~\cite{latoza2014microtask}.\par

In this paper, we investigate approaches for offering immediate feedback within a microtask. Developers receive feedback by executing unit tests, by syntax errors, and by observing the execution of the program through debugging.

\subsection{Fast onboarding}
Another challenge in using crowd work within a software project is the process of onboarding. A number of studies have documented the joining scripts used and barriers that open source software developers face when onboarding onto a new project. These include installing necessary tools, downloading code from a server, identifying and downloading dependencies, and configuring the build environment \cite{STEINMACHER2015,joiningOSS_Alex2003,Onion_Patch_Jergensen_2011,fagerholm2014onboarding}. As a result, onboarding onto open source projects can require weeks of time, creating a substantial barrier dissuading casual contributors from joining. 

Researchers have explored programming environments which aim to alleviate these barriers. 
Codepilot \cite{CodePilot:2017} reduces the complexity of programming environments for novice programmers
by integrating a preconfigured environment for real-time collaborative programming, testing, bug reporting, and version control into a single, simplified system.
In Collabode, multiple developers synchronously edit code at the same time, enabling new forms of collaborative programming \cite{Collabode:2011,Collabode:2012}. 
Apparition offers an online environment for building UI mockups, offering an integrated environment for authoring, viewing, and collaborating on the visual look and feel and behavior of UI elements~\cite{lasecki2015apparition}. CrowdCode offers an online preconfigured environment for implementing libraries, enabling developers to onboard quickly onto programming tasks \cite{latoza2018microtask}. 

In this work, we build on these approaches, offering a preconfigured environment for fast onboarding specifically designed for implementing microservices. 

\subsection{Behavior Driven Development (BDD)}

In this paper, we apply behavior-driven development to microtask programming. BDD focuses on defining fine-grained specifications of a system's behavior in a way that they can be tested ~\cite{smart2014bddInActionBook}. This enables writing executable specifications of a system ~\cite{gomez2018analysis}. An acceptance test in BDD is a specification of the system's behavior that verifies its behavior rather than its state. A survey of literature and current BDD toolkits identified several characteristics of BDD, including ubiquitous language, iterative decomposition, plain text descriptions of user stories and scenario templates, automated acceptance testing with mapping rules, and readable behavior-oriented specification code ~\cite{gomez2018analysis}. 

There are few studies investigating the impact of applying BDD. One reason may be that the original version of BDD is highly similar to test-driven development. Several proponents believe BDD helps teams to generate and deliver higher quality software quickly~\cite{smart2014bddInActionBook, BDDWEbsite, north2006introducing,rahman2015reusable,smart2014bddInActionBook}. One study of 22 developers analyzed BDD's impact on the software life cycle [13]. The study found that BDD increased the quality of software products by 2\% and 5\% in relation to TDD and traditional iterative tests, respectively~\cite{gomez2018analysis}.

 \begin{figure}
\includegraphics[width=\columnwidth,keepaspectratio, clip]{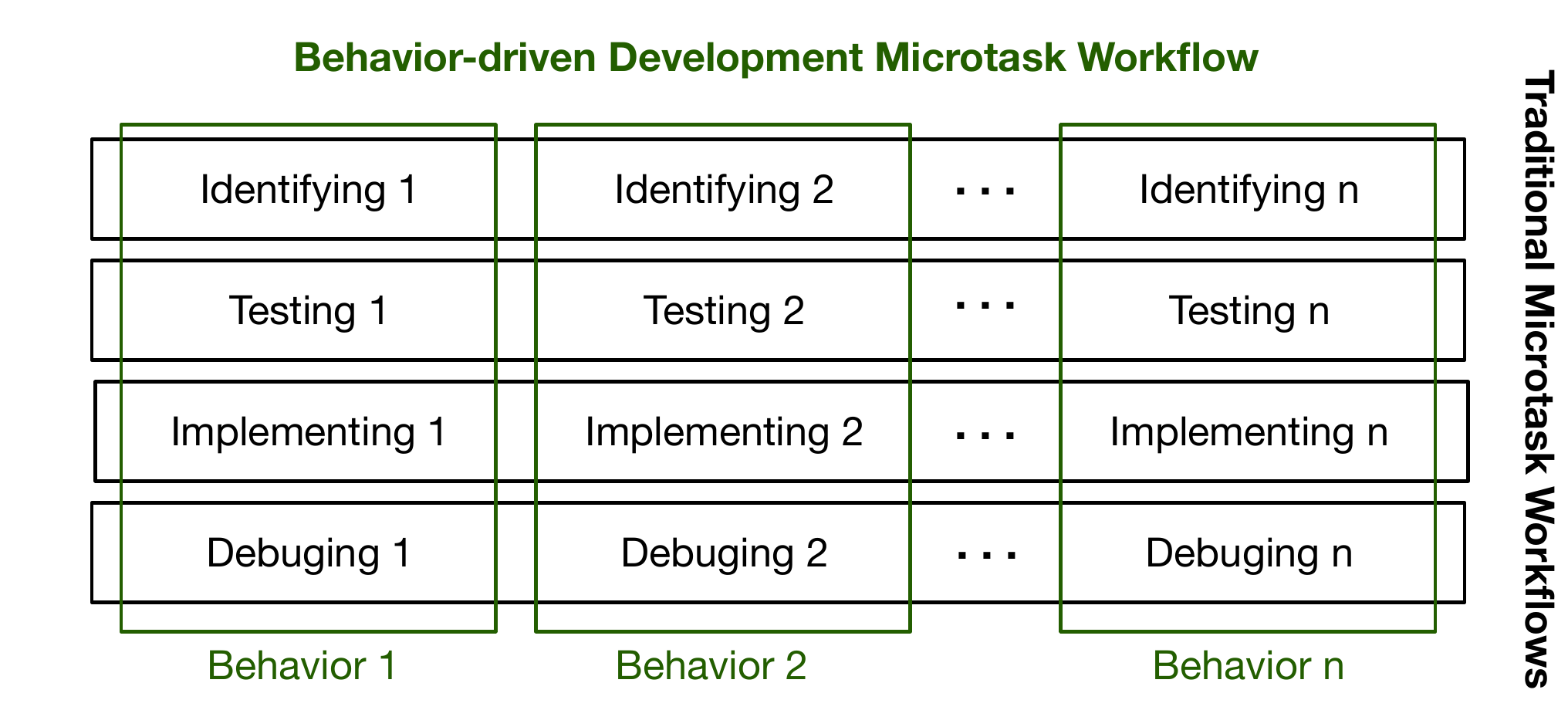}
\centering
\caption{Traditional microtask programming decomposes large programming tasks into separate microtasks in which different types of contributions can be made, such as identifying all of the behaviors which should be tested in a function or writing a test. In behavior-driven microtask programming, work is instead decomposed by behavior, where an individual microtask incorporates work of several types for an individual behavior. }
  ~\label{fig:workflow}
\end{figure}

\section{Crowd Microservices workflow}

In this section, we describe a behavior-driven approach to microtask programming. The main goal of the workflow is to reduce onboarding time and increase the potential for parallelism. In behavior-driven development, developers first write a unit test for each behavior they will implement, offering a way to verify that their implementation works as intended by running a test. As a workflow for microtasking, behavior-driven development offers a number of potential advantages. As developers work, they receive feedback before submitting, enabling the developer to revise their own work. Rather than requiring separate developers to test, implement, and debug a function through separate microtasks and coordinate this work to ensure consistency, a single contributor focuses on work related to an individual behavior within a function (Fig. \ref{fig:workflow}). 
\par

We apply our approach to implementing microservices. Web application back-ends are often decomposed into numerous microservices, offering a natural boundary for crowdsourcing a module that is part of a larger system. In our approach, a client, for example a software development team, may choose to crowdsource the creation of an individual microservice. In situations where teams lack sufficient developer resources to complete work sufficiently quickly, a development team might choose to delegate this work to a crowd. A client, acting on behalf of the software development team, may define the desired behavior of the microservice by defining a set of endpoints.\par

\begin{figure*}
\centering
\includegraphics[width=\textwidth]{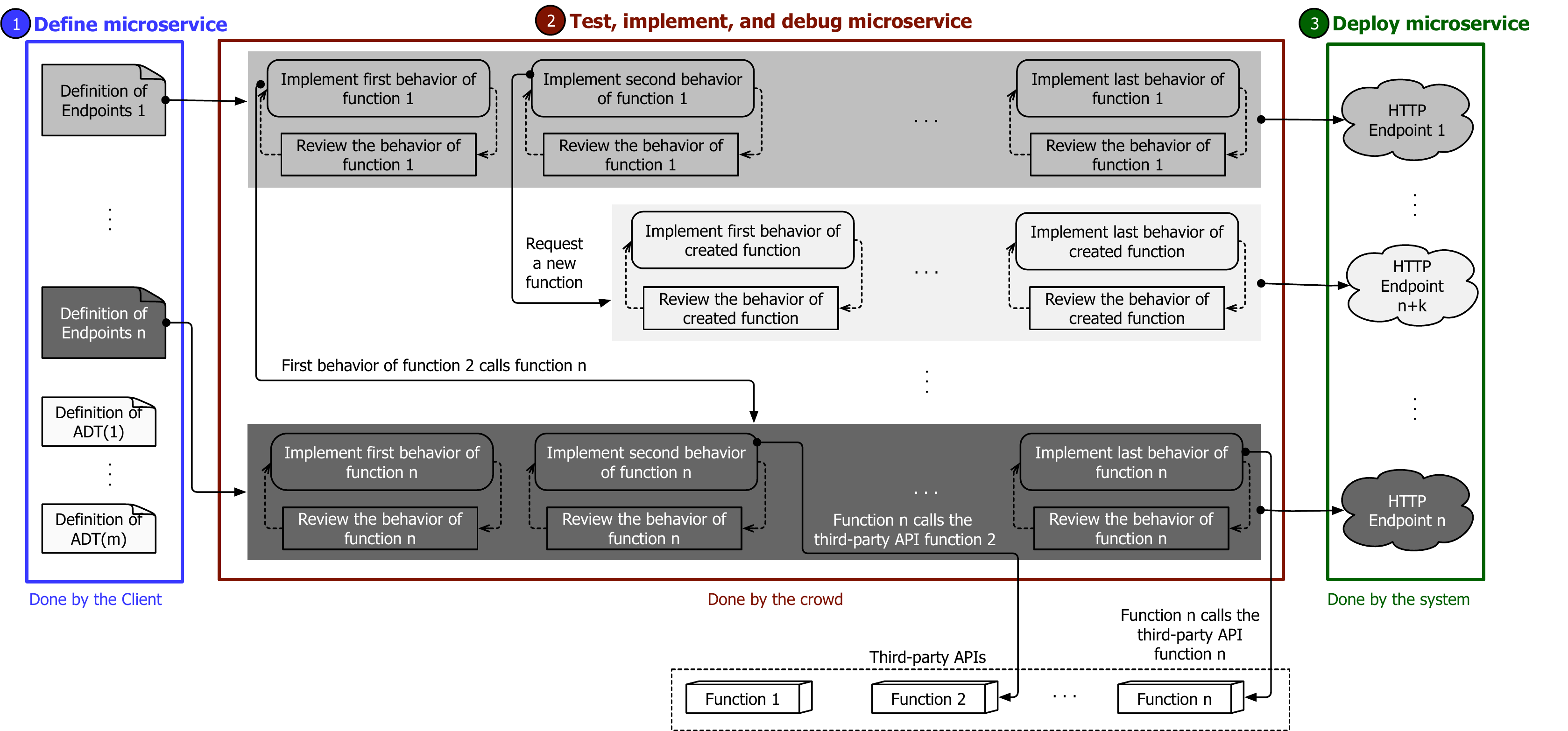}
\caption{ \textcolor{myBlue}{\textbf{(1) Define microservices}}: The client first writes a \texttt{Client-Request} to define a microservice to implement.\textcolor{Maroon}{\textbf{(2) Test, implement, and debug microservice}}: The system generates microtasks as necessary to implement each endpoint. \textcolor{OliveGreen}{\textbf{ (3) Deploy microservice}}: After implementation is complete, the client may deploy the microservice.}
\label{fig:systemArchitecture}
\end{figure*}

In the following sections, we describe our behavior-driven workflow and its application to implementing a microservice. Fig.~\ref{fig:systemArchitecture} surveys our approach.

\subsection{Microtasks}

In our behavior-driven microtask workflow, contributions are made through two microtasks: \texttt{Implement Function Behavior} and \texttt{Review}. 
Table.~\ref{table:microtasksType} summarizes the context and possible contributions of each. \par
 \begin{table*}

\caption{\textit{Crowd Microservices} enables workers to make contributions through two microtask types. Each offers editors for creating content, context views that make the task self-contained by offering necessary background information, and contributions the worker may make.}
    \centering
    \rowcolors{1}{gray!40}{}
    \begin{tabular}{p{2.8cm}p{3.8cm}p{5cm}p{4cm}}
        \textbf{Microtask types}   &  \textbf{Editor} & \textbf{Context view} & \textbf{Possible contributions} \\
        \hline

        \textbf{Implement Function Behavior} & 1) Function and test editor, 2) test runner, 3) stub editor & 1) Description and signature of function 2) description of requesting function 3) ADTs & 1) Implement behavior 2) Report an issue in function 3) Mark function as completed  \\
        \textbf{Review} &Rating and review text & 1) Description and signature of function 2) implementation of function 3) function unit tests 4) ADTs &  Rating and review \\
    \end{tabular}
\label{table:microtasksType}
\end{table*}
\subsubsection{Interacting with microtasks}
After logging in, workers are first taken to a welcome page which includes a demo video and a tutorial describing basic concepts in the \textit{Crowd Microservices} environment. After completing the tutorial, workers are taken to a dashboard page, which includes the client's project description, a list of descriptions for each function, and the currently available microtasks. The system automatically assigns workers a random microtask, which the worker can complete and submit or skip. When workers begin a type of microtask which they have not previously worked on, workers are given an additional tutorial explaining the microtask. When participants work and become confused about a design decision to be made or about the environment itself, they may use the global~\texttt{Question and Answer} feature to post a question, modeled on the question and answer feature in CrowdCode~\cite{LaToza2015Scaffolding}. Posted questions are visible to all workers, who may post answers and view previous questions and answers. Each project defined in the \texttt{Client-Request} (Figure.\ref{fig:Client_Request_subsystem}) has its own ~\texttt{Question and Answer}. \par

As workers complete microtasks, each contribution is given a rating through a \texttt{Review} microtask. Ratings are then summed to generate a score for each worker. This score is visible to the entire crowd that participated on the project on a global \texttt{leaderboard}, helping to motivate contributions and higher quality work. As workers probably watched the scores of others, they might be motivated to increase their score and increase their ranking above other workers. Crowd workers could achieve higher scores by submitting more either \texttt{Review} or \texttt{Implement Function Behaviors}, submitting each \texttt{Review} task worth five scores for the reviewer. Submitting an \texttt{Implement Function Behavior} gets a score based on the number of rating stars it received by a review. Each star is worth 2 points. Consequently, each \texttt{Implementation Function Behavior} based on its quality could collect score 2 to 10 scores. 
\par

\subsubsection{Implement Function Behavior microtask}
Workers perform each step in the \texttt{Implement Function Behavior} Microtask through the \textit{Crowd Microservices}  environment  (Fig.\ref{fig:Implement_Fuction_Behavior_all}). 
\par 
\begin{enumerate}[Step 1.]
\item Identify a Behavior.
Workers first work to identify single behavior that is not yet implemented from the comments describing the function located above its body (see (1) in Fig.  \ref{fig:Implement_Fuction_Behavior_all}). When the function has been completely implemented and no behaviors remain, the worker may indicate this through a checkbox. \par

\item Test the Behavior.
In the second step, workers author a test as a simple input/output pair, specifying inputs and an output for the function under test, or as an assertion-based test (see (2) in Fig. \ref{fig:Implement_Fuction_Behavior_all}). The worker may run the test to verify that it fails with the current implementation. In the test editor worker can write assertion unit tests to evaluate the behavior. Also, in the test editor worker can invoke third-party APIs or the other functions. \par

\item Implement the Behavior.
The worker next implements their behavior using a code editor (see (3) in Fig.\ref{fig:Implement_Fuction_Behavior_all}). When the behavior to be implemented is complex, the worker may choose to create a new function, specifying a description, name, parameters, and return type. They may then call the new function in the body of their main function. After the microtask is submitted, this new function will be created, and a microtask generated to begin work on the function. In some cases, the signature of the function that a worker is asked to implement may not match its intended purpose, such as missing a necessary parameter. In these cases, the worker cannot directly fix the issue, as they do not have access to the source code of each call site for the function. Instead, they may report an issue, halting further work on the function. The client is able to see that this issue has been created, resolve the problem, and begin work again.  
\par

\item Debug the Behavior.
A worker may test their implementation by running the function's unit tests. As it is shown in Fig. \ref{fig:debug} when a test fails, workers may debug by using the \texttt{Inspect code} feature to view the value of any expression in the currently selected test. Hovering over an expression in the code invokes a popup listing all values held by the expression during execution. In cases where a function that is called from the function under test has not yet been implemented, any tests exercising this functionality will fail. To enable the worker to still test their contribution, a worker may create a stub for any function call. Creating a stub replaces an actual output (e.g., an undefined return value generated by a function that does not yet exist) with an intended output. Using the \texttt{stub} editor (Fig. \ref{fig:stubEditor}), the worker can view the current output generated by a function call and edit this value to the intended output. This then automatically generates a stub. Whenever the tests are run, all stubs are applied, intercepting function calls and replacing them with stubbed values.  \par

In pilot studies, we found that requiring workers to only submit microtasks that did not contain errors decreased the productivity of workers dramatically. In those cases, contributors spent the entire 15 minutes, the maximum, before being forced to skip the microtask, losing their work. We therefore enable workers to submit incomplete work and get feedback on incomplete implementations from reviewers. 
 \par

\item Submit the microtask.
Once finished, the worker may submit their work. To ensure that workers do not lock access to an artifact for extended periods of time, each microtask has a maximum time limit of 15 minutes.
We derived the 15 minutes from our previous research \cite{latoza2018microtask}, 
where the average time for each microtask was less than 5 minutes. 
This constraint helps the crowd to focus on their microtask to complete or skip it in less than 15 minutes. Workers are informed by the system when time is close to expiring. When the time has expired, the system informs the worker and skips the microtask.
\end{enumerate}

\subsubsection{Review microtask}
In the \texttt{Review} microtask, workers assess contributions submitted by other workers. Workers are given a diff comparing the code they submitted with the previous version as well as the tests of the function. Instead of being asked to make a binary choice to accept or reject the contribution, workers are asked to assign a rating of 1 to 5 stars. If the worker evaluates the work with 1 to 3 stars, the work is marked as needing revision. The worker then describes aspects of the contribution that they feel need improvement, and a microtask is generated to do this work. If the worker evaluates the submitted contribution with 4 or 5 stars, the contribution is accepted as is. In this case, the assessment of the work is optional, which will be provided back to the crowd worker that made the contribution. When a contributor is notified that his or her contribution is accepted but it did not receive full stars, the workers receives a notification with feedback which they may use for increasing the quality of their future contributions. 

\subsection{Assembling microservices}
\begin{figure*}
\includegraphics[width=\textwidth,keepaspectratio,clip]{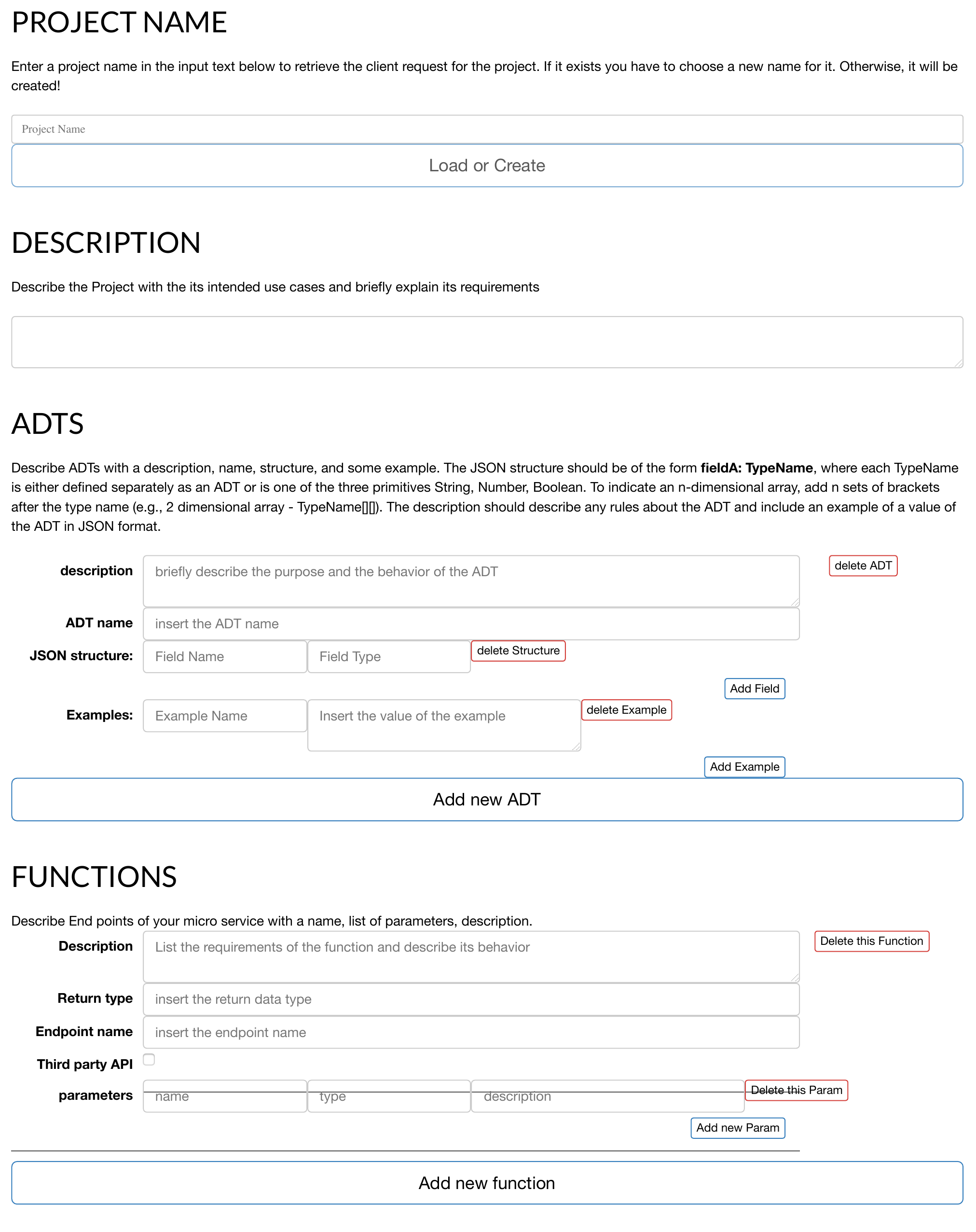}
\centering
\caption{In the \texttt{Client-Request}, a client defines a microservice they would like created by describing a set of endpoints and corresponding data structures (ADTs). }
   \label{fig:Client_Request_subsystem}
\end{figure*}

Our approach applies the behavior-driven development workflow to implementing microservices. Fig. \ref{fig:systemArchitecture} depicts the steps in our process. The microservice is first described by a client through the \texttt{Client-Request} page (Fig \ref{fig:Client_Request_subsystem}). Clients define a set of endpoints describing HTTP requests which will be handled by the microservice. Each endpoint is defined as a function, specifying an identifier, parameters, and a description of its behavior. As endpoints may accept complex JSON data structures as input and generate complex JSON data structures as output, clients may also describe a set of abstract data types (ADTs). Each ADT describes a set of fields for a JSON object, assigning each field a type which may be either a primitive or another ADT. In defining endpoints, clients may specify the expected data by giving each parameter and return value a type. \par

After a client has completed a~\texttt{Client-Request}, they may then submit this \texttt{Client-Request} to generate a new \textit{Crowd Microservices} project. As shown in step 2 of Fig. \ref{fig:systemArchitecture}, submitting a~\texttt{Client-Request} generates an initial set of microtasks, generating an \texttt{Implement Function Behavior} microtask for each endpoint function. Workers may then log into the project to begin completing microtasks. As workers complete microtasks, additional microtasks are automatically generated to review contributions, continue work on each function, and implement any new functions requested by crowd workers. \par

Microservices often depend on external services exposed through third-party APIs. As identifying, downloading, and configuring these dependencies can serve as a barrier to contributing, \textit{Crowd Microservices} offers a pre-configured environment. As typical microservices often involve persisting data between requests, we chose to offer a simplified API for interacting with a persistence store. 
Through this API, workers can store, update, and delete JSON objects in a persistence store. Workers may use any of these API functions when working with functions and unit tests in the \texttt{Implement Function Behavior}  microtask. Any schema-less persistence store may be used as an implementation for this API. In our prototype IDE, a development version used by workers simulates the behavior of a persistence store within the browser and clears the persistence store after every test execution. 
In the production version used after the microservice is deployed, the API is implemented through a Firebase store.\par


After the crowd finishes the implementation of a microservice, the  client may choose to create and deploy the microservice to a hosting site (step 3 in Fig. \ref{fig:systemArchitecture}). Invoking the \texttt{Publish} command first creates a new node.js GitHub project which includes each function implemented by the crowd. For endpoint functions, the environment automatically generates an HTTP request handler function for the endpoint. An example is shown in Fig.\ref{fig:microSample}. In this example, a signature \texttt{function fetchTodosBasedOnStatus(userId, status)} would generate a GET \texttt{/fetchTodosBasedOnStatus} endpoint with the parameters as fields in the body of the request. Each endpoint then contains the implementation of the function defined by the crowd. 
Next, this GitHub project is deployed to a hosting site (Heroku in our prototype implementation). A new project instance is created, and the project is deployed. After this process has completed, the client  may begin using the completed microservice by making HTTP requests against the deployed, publicly available microservice. As some projects may require private rather than public deployment, the client may also provide information for a private repository to deploy to through the \texttt{Client-Request}. 


\begin{figure}
\includegraphics[width=\columnwidth,keepaspectratio,clip]{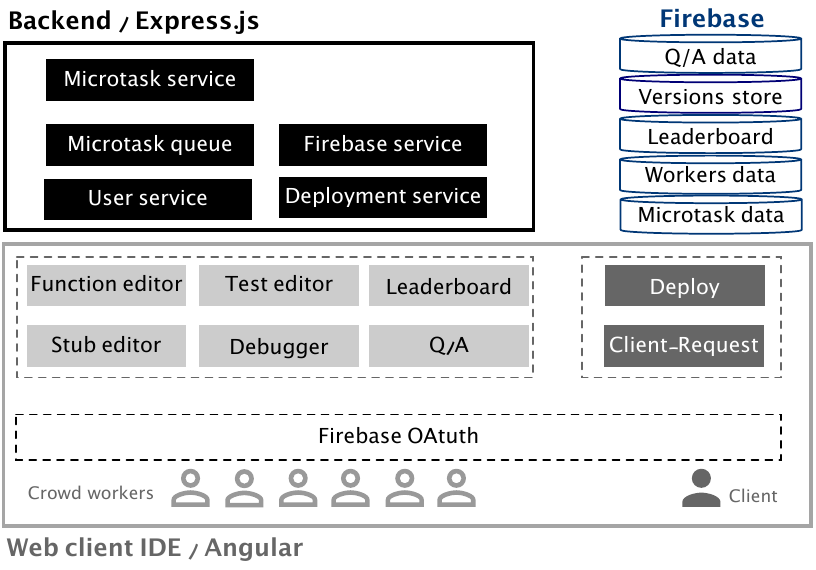}
\centering
\caption{Crowd Microservices is implemented as a client-server application with a backend, a real-time datastore, and web clients. }
   \label{fig:architecture_Crowd_microservices}
\end{figure}

\subsection{Implementation}
We implemented our approach as a prototype \textit{Crowd Microservices} IDE~\footnote{https://youtu.be/qQeYOsRaxHc}. As shown in  Figure.~\ref{fig:architecture_Crowd_microservices}, \textit{Crowd Microservices} is a client-server application with three layers: 1) a web client, implemented in AngularJS, which runs in a worker's browser, 2) a back-end, implemented in Express.js, and 3) a persistence store, implemented using the Firebase Real-time Database \footnote{https://firebase.google.com}. 
\textit{Crowd Microservices} automatically generates microtasks based on the current state of submitted work. After a \texttt{Client-Request} defines endpoints, the system automatically generates a function and microtask to begin work on each. 
After an \texttt{Implement Function Behavior} microtask is submitted, the system automatically creates a \texttt{Review} microtask. After a~\texttt{Review} microtask is submitted, an \texttt{Implement Function Behavior} is generated to continue work, if the contributor has not indicated that work is complete. If a review of an \texttt{Implement Function Behavior} contribution indicates issues that need to be fixed, a new \texttt{Implement Function Behavior} microtask is generated, which includes the issue and an instruction to fix it. After a microtask is generated, it is added to a queue. When a worker fetches a microtask, the system automatically assigns the worker the next microtask and removes it from the queue.

\begin{figure*}
\centering
\includegraphics[width=\textwidth,keepaspectratio, clip]{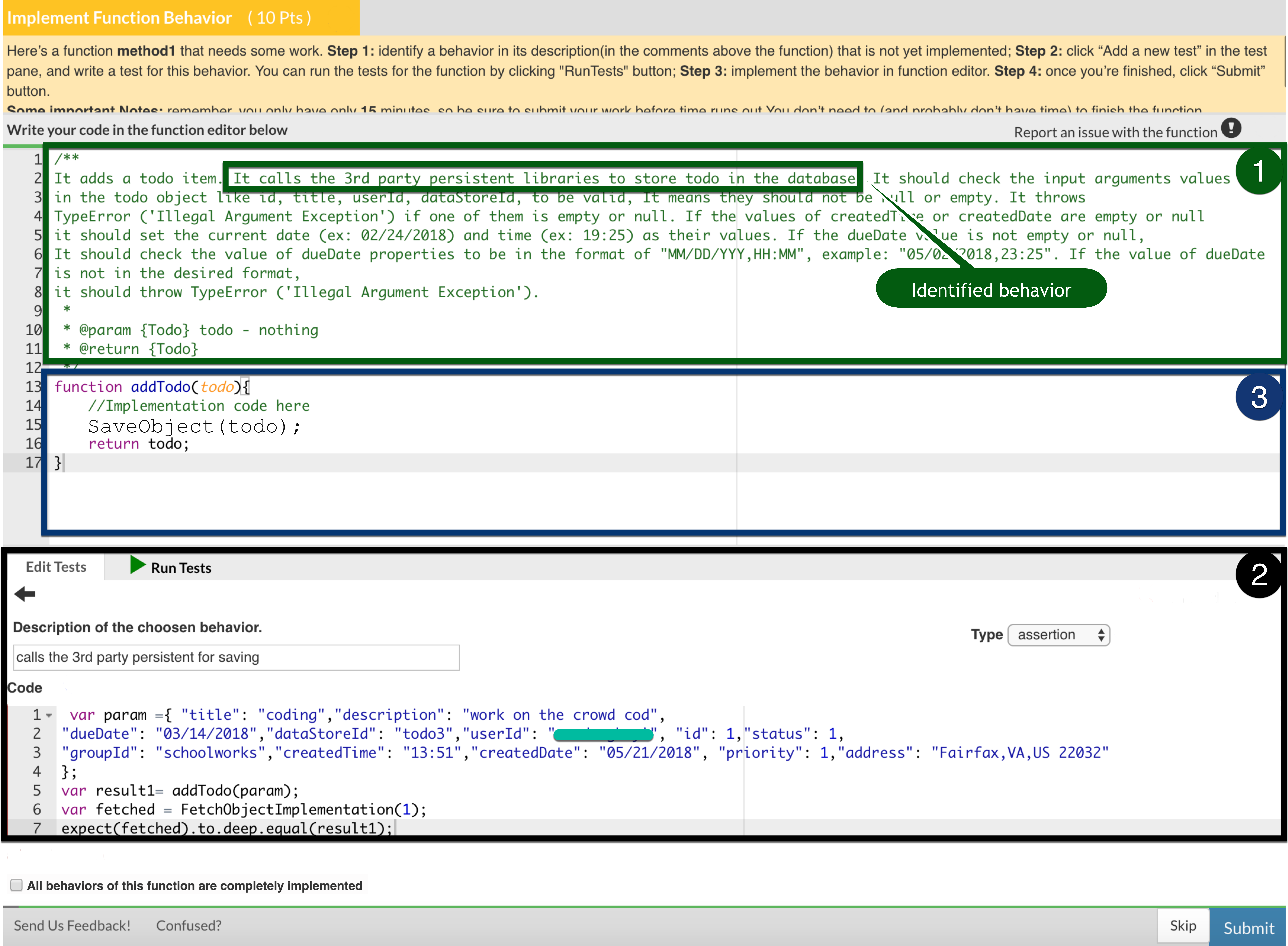}

\caption{In the \texttt{Implement Function Behavior} microtask, workers first \textcolor{OliveGreen}{\textbf{(1) identify a behavior from the description of a function.}} They then \textcolor{Black}{\textbf{(2) write a test in the test editor to verify the behavior}}. and \textcolor{RoyalPurple}{\textbf{(3) edit the code for the function to implement it}}. Finally, they (4) test it by running the tests, fixing issues they identify. }
  \label{fig:Implement_Fuction_Behavior_all}
\end{figure*}

\section{Example}
We illustrate using \textit{Crowd Microservices} to build a microservice through an example. A client first requests a microservice (Section \ref{Example1}), crowd workers implement behaviors, review contributions, and create new functions (Section \ref{Example2}), and the client may then deploy the microservice (Section \ref{Example3}).

\subsection{The client requests a microservice}\label{Example1}
Bob decides that he would like to add a \textit{ToDo} widget to his app. Hoping to add it as soon as possible, he decides to create the UI himself and to crowdsource implementing the back-end through \textit{Crowd Microservices}. Using the \texttt{Client-Request} page (Fig. \ref{fig:Client_Request_subsystem}),
he creates a new ToDo microservice project, providing a brief description, defining a set of endpoints, and describing the format of the JSON data used in the request and responses through the data structures editor. He then goes back to implementing the UI while the crowd begins implementing the back-end. \par

\subsection{The crowd develops  the microservice}\label{Example2}
\subsubsection{Implementing behaviors}
Alice, the first contributor, logs in to \textit{Crowd Microservices}, notices the ToDo microservice project has a number of open microtasks, and decides to contribute. She watches a short video
\footnote{https://youtu.be/mIn2EOqsDYw}
and reads a short tutorial to familiarize herself with the environment. Viewing the dashboard for the project, she quickly reads the brief description from the client and a list of descriptions of the functions created by the crowd so far.  She clicks \texttt{Fetch a Microtask}, and \textit{Crowd Microservices} assigns her an \texttt{Implement Function Behavior} microtask (Figure \ref{fig:Implement_Fuction_Behavior_all}). Following the behavior-driven workflow, she first reads the description of the function in the comments above the body, identifying a behavior that does not yet seem to be implemented.  Next, she writes a traditional unit test, writing test code with assertions and running the test to verify that it fails. Next, Alice implements the behavior using the function editor. Alice tests her implementation, running the function's tests, but one fails.  To understand why, Alice using the debugger to inspect values in the function's execution, hovering over several variables to see their values. Identifying the issue, Alice fixes the problem, finds that all of the tests now pass, and submits. \par

As Alice works, the crowd simultaneously completes other microtasks. After logging in, Dave is assigned an \texttt{Implement Function Behavior} microtask. Unsure how to implement any of the remaining behaviors, he clicks \texttt{Skip} and fetches another microtask. This time, he thinks that he can complete this microtask and writes a test and implementation. After a test fails, he realizes he does not know how to correctly call a third-party API function. Using the \texttt{Question and Answer} feature,  he asks, \textquote{How  can I store a todo object in the database?} A worker responds, and he figures out how to fix the problem. However, he then sees an alert: \textquote{You have spent more than 14 minutes on the current microtask, so try to submit your task in one minute before the system automatically skips it}. Wanting to submit his partially completed work, Dave submits with failed tests. Fetching another microtask and inspecting the implementation, he see that that all behaviors have already been implemented. So he clicks the corresponding checkbox and submits.\par

\subsubsection{Reviewing contributions}
After logging in, Oliver is assigned a \texttt{Review} microtask which asks him to assess the behavior implemented by Alice. Oliver reads the description of the function, a diff of the code written by Alice, and the tests. Looking at the implementation, he finds it seems incomplete, so he rates her contribution a 2 on a 5 point scale and gives her feedback, \textquote{The behavior asked you to evaluate all input arguments of the function, but you just checked the validity of the date.} Oliver submits, and Alice receives a notification that her work was reviewed and received 2 stars. \par
\begin{figure*}
\includegraphics[width=\textwidth,keepaspectratio, clip]{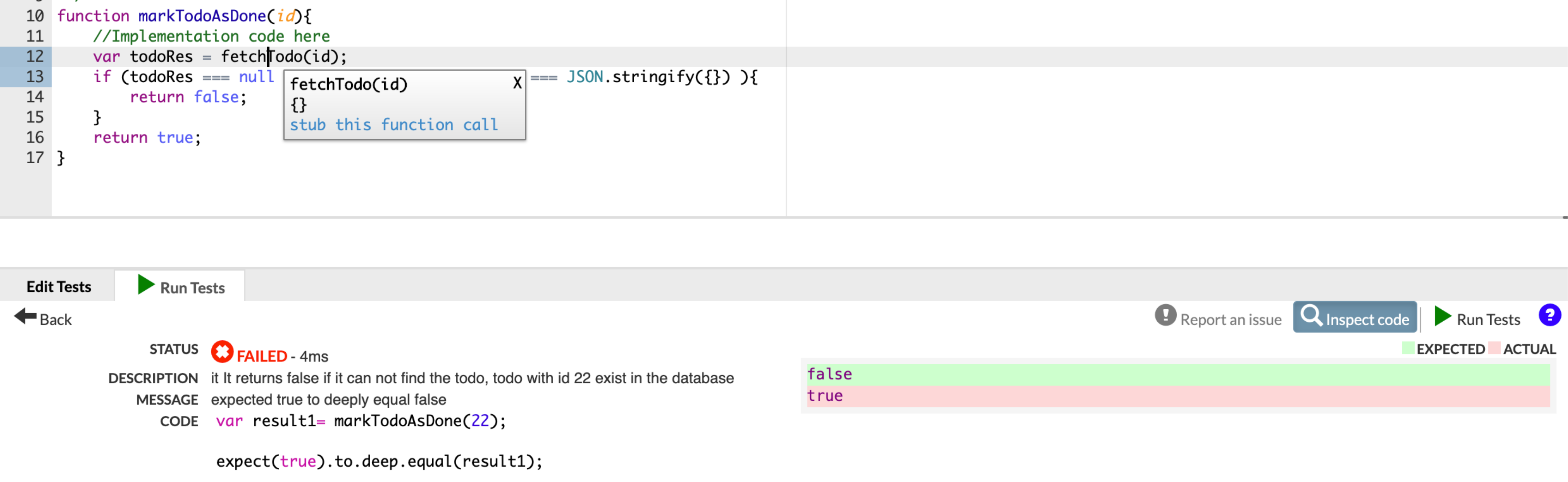}
\centering
\caption{Workers can debug failed tests by by using the \texttt{Inspect code} feature to view the value of any expression. In cases where a function that is called from the function has not yet been implemented, any tests exercising this functionality will fail. To enable the worker to still test their contribution, a worker may create a stub for any function call.}
  ~\label{fig:debug}
\end{figure*}

\subsubsection{Creating a new function}
\begin{figure}
\includegraphics[width=\columnwidth,keepaspectratio,clip]{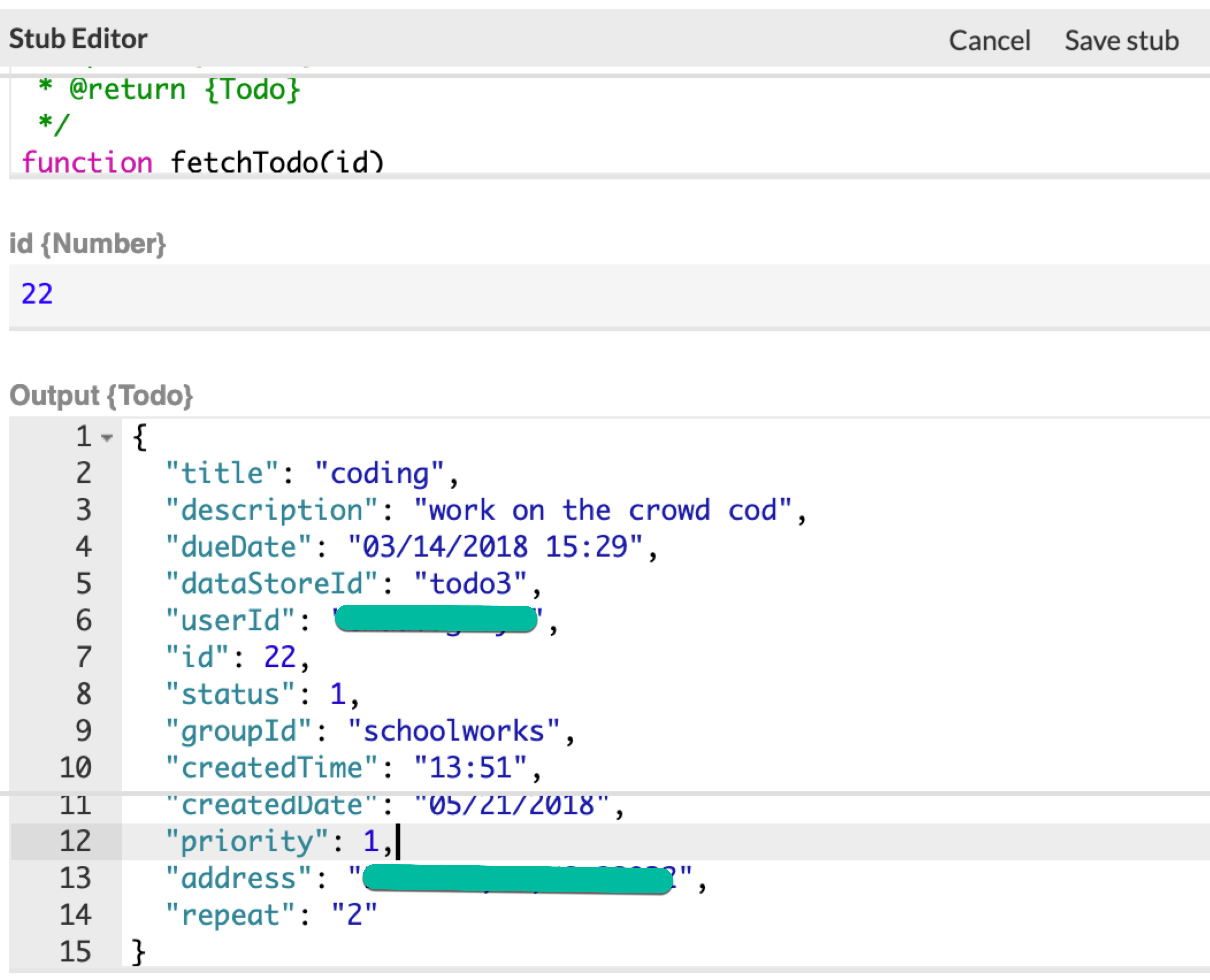}
\centering
\caption{Workers can view the current output generated by a function call and edit this value to the intended output. This then automatically generates a stub with the new value, which is applied when the tests are run. }~\label{fig:stubEditor}
\end{figure}

After being assigned an \texttt{Implement Function Behavior} microtask, Jon decides to implement a format check for the  \texttt{todoDate} parameter. Believing this to be fairly complex, he decides it would be best implemented in a separate function. He invokes the \texttt{create a new function} feature, creating the function \texttt{checkTodoDataFormat} for others to implement. Specifying its behavior, he writes a description and signature. He then calls this new, currently empty, function from the body of the function he is working on. To verify his work, he runs the tests. But as \texttt{checkTodoDateFormat} is not yet implemented, his tests fail (Fig. \ref{fig:debug}). Jon uses the \texttt{stub} editor (Fig. \ref{fig:stubEditor}) to replace the actual output with a stub value representing the desired output. This automatically replaces all calls to this function with the inputs and output Jon specifies. Jon runs the tests again, they pass, and he submits. \par

As the crowd works, each is assigned a score  based on the ratings of their contributions. These scores are visible on a global \texttt{leaderboard} visible to the project's entire crowd, encouraging everyone to work hard to place higher.\par

\subsection{The client deploys the microservice}\label{Example3}
While the crowd was working, Bob implemented the front-end, inserting requests based on the behavior of the endpoints he specified. 
After all the microtasks have been completed and the implementation finished, he clicks a button to deploy the microservice. He loads his web app in his browser, seeing the \textit{ToDo} interactions handled by the microservice. \par


\section{Evaluation}
To investigate the feasibility of applying behavior-driven microtask programming to implementing microservices, we conducted a user study in which a crowd of workers built a small microservice.
Since there are no prior systems for implementing microservices through microtasks, we conducted a study evaluating the feasibility of the system rather than comparing it against existing approaches. A key goal of microtask programming is to enable short contributions by transient contributors. Therefore, we evaluated how long it takes for a new crowd worker to onboard and make a contribution. 
Specifically, we investigated
(1) the feasibility of crowd workers to make contributions through a behavior-driven microtask workflow,
(2) the time necessary to onboard and make a contribution,
(3) the feasibility of implementing and testing a microservice entirely through microtasks.

We recruited 9 participants to build a small microservice for a \textit{ToDo} application and then analyzed their environment interactions and the resulting code they created. 
\subsection{Method}
We recruited nine participants by advertising on Facebook, LinkedIn, and Twitter and through flyers (referred to as P1-P9).  Participants connected to our system from the US, Spain, England, and India. Each had prior experience in JavaScript. Participants included one undergraduate student in computer science or a related field (P5),  one instructor (P9), and seven graduate students in computer science or related fields. As typical in open contribution platforms, participants exhibited a diverse range of experience, with prior experience in JavaScript including less than 6 months (P1 and P2), 7-12 months (P3, P4, and P5), and more than 4 years (P6, P7,P8, and P9).\par

We split our study into two sessions to reduce participant fatigue as well as to simulate participants returning to work after a delay, as is common in microtask work.  All nine participants participated in the first session and five (P1, P5, P6, P8, and P9) participated in the second session. The first session was 150 minutes, and the second 120 minutes. One participant (P8) left the second session early after approximately one hour.
All worked entirely online at their own computers, and their interactions with other participants were only via the \texttt{Question and Answer} feature. Participants were paid 20 dollars per hour for their time through gift cards.\par

The crowd worked to build a microservice for the back-end functionality of a \textit{ToDo} app. We first examined several \textit{ToDo} applications to identify common functionality. We then wrote a client request containing 12 endpoints which asked developers to implement functionality for creating, deleting, updating, fetching, reminding, and archiving todo items.\par

We gathered data from several sources. Before beginning the study, participants completed a short demographics survey. During the study session, all participant interactions with \textit{Crowd Microservices} were logged with a timestamp and participant id. This included each microtask generated, submitted, and skipped as well as each change to a function or test. 
At the end of the first session, participants completed a survey on their experiences with \textit{Crowd Microservices}, focusing specifically on their experience with the behavior-driven development workflow. At the end of the study, five participants participated in a short 15 minute semi-structured interview. The open-ended questions focused on onboarding challenges, the granularity of microtasks, the ability to choose a task, motivation working using microtask programming, and interactions between crowd workers.\par

 \begin{table*}
\caption{Number of microtask completions, skips, and completion times.}
\rowcolors{2}{}{gray!40}
\begin{tabular}{p{5cm}p{1.5cm}p{1.5cm}p{1.5cm}p{1.5cm}p{1.5cm}p{1.5cm}}
        &\multicolumn{2}{c}{\textbf{Completed}} & \multicolumn{2}{c}{\textbf{Skipped} }& \multicolumn{2}{c}{\textbf{Median time (mm:ss)} } \\

           \textbf{Microtasks types}  & Session 1&Session 2 &Session 1&Session 2&Session 1& Session 2\\
        \hline
        \textbf{Implement Function Behavior} & 112 &63 & 39 & 22 & 4:12 & 3:27 \\

        \textbf{Review} & 112 &63 & 5 & 9 & 2:41 &  2:25  \\
        
      \textbf{Total} &  \multicolumn{2}{c}{350} & \multicolumn{2}{c}{75} &\multicolumn{2}{c}{--}\\
    \end{tabular}

    \label{table:microtasksCompletedSkipped}
\end{table*}

At the beginning of the study, participants logged in to \textit{Crowd Microservices} and worked through tutorial content,  watching a tutorial video, reading the welcome page, and reading a series of 6 tutorials on using the individual microtasks. Participants then began work by fetching a microtask.  Participants were allowed to use Internet searches as they saw fit. 

The study replication package is publicly available~\footnote{https://github.com/devuxd/crowd-microservices-output/tree/master/replication-package/todo}. It includes the study materials, client requests, and test suite used to evaluate contributions as well as the code written by the crowd in our study.

\subsection{Results}

\subsubsection{Feasibility of behavior-driven microtasks}
To investigate the ability of participants to use the behavior-driven microtask workflow, we examined the log data to determine how many microtasks participants were able to successfully complete during the two sessions as well as the functions and tests they created. Overall, participants successfully submitted 350 microtasks and implemented 13 functions, one of which was defined by the crowd (Table~\ref{table:microtasksCompletedSkipped}). Participants created a test suite of 36 unit tests, writing an average of 3 unit tests per function. We analyzed the number of lines of code in each function and test, counting the final numbers of lines in each at the end of the study. 
Participants wrote 216 lines of code, approximately 16 lines per function. Participants wrote 397 line of code in their test suite. 


Several participants reported that identifying a behavior was not difficult. Three participants reported that they preferred to focus on easy behaviors first:
\begin{quote}"I chose the easiest behavior to implement first. This was usually to check if the input was null or empty. If that was already implemented, I just went in order." - (P5)\end{quote}
Others reported that some behaviors were not clear, leading them to focus first on those which were:
 \begin{quote}"I chose based on being more clear and simple to me. Sometimes it wasn't clear what exactly that behavior means." - (P8) \end{quote}

The interview and survey results revealed positive impacts of the \texttt{leaderboard} on crowd workers. Firstly, two participants in the interview shared since there was a direct relationship between the quality of contributions and scores that reviewers gave to those contributions, we tried to submit our task with higher quality. Crowd workers tried to submit microtasks that be accepted by reviewers and receive 8 or 10 scores. Secondly, workers reported that it motivated them to achieve higher scores to achieve a higher ranking. \par
\begin{quote}"The leaderboard was good for team KPIs [key performance indicator] and healthy competition. " - (P4) \end{quote}

The acceptance rate of \texttt{Implementation Function Behavior} tasks by reviewers was lower in the second session than in the first. 
In the first session 85\% were accepted by reviewers while in the second session 40\% were accepted by reviewers. 
This was also reflected in the leaderboard scores. The average scores in the leaderboard were 129.5 points for the first session and 103.2 for the second, with minimum and maximum scores of 54 and 241 and 13 and 151 (the participant with the score of 13 contributed only two microtasks before experiencing technical issues).
Accepted \texttt{Implementation Function Behavior} contributions were worth 8 to 10 points, rejected contributions were worth 2 to 6 points, and \texttt{Review} tasks were worth 5 points.

There are several potential explanations as to why acceptance rates were lower in the second session. In the second session, participants continued work on incomplete behaviors left from the first session which may have been harder to complete. Survey data revealed that 3 participants preferred writing easier behaviors first and harder behaviors later. For example, easier behaviors like calling a third-party API , \textquote{It calls the 3rd party persistent libraries to store todo in the database}, were implemented sooner in the first session (Figure~\ref{fig:Implement_Fuction_Behavior_all}, green box 1).  Harder behaviors were implemented in the second session, such as  \textquote{It should check the value of \texttt{dueDate} properties to be in the format of 'MM/DD/YY,HH:MM'. If the value of \texttt{dueDate} is not in the desired format, it should throw \texttt{TypeError ('Illegal Argument Exception')}}. 


\par

As participants were not strictly required by the workflow to test first, two participants in the post-task survey shared with the experimenters that they chose to first implement the function body and then implement the test for the function body.\par
\begin{quote}"I usually implemented the body first just so I could figure out how the function should work."-(P5)\end{quote} 
\begin{quote}"I would interchange step 2 and 3 so that the user first implements and then tests the behavior." - (P7)\end{quote} 

The workflow offered crowd workers the freedom to decompose a complex function into two or more functions. 
Participants implemented 13 functions, including creating one function. P8 created a function \texttt{checkDateFormat}. As it was not yet implemented by the other participants, he used the \texttt{stub} feature to simulate its behavior.\par

One participant, confused about the creation of a todo object, asked a question using the \texttt{Question and Answer} feature. Other participants responded to this question. Another participant asked a question about a function description, to which the other participants replied.\par

Throughout both sessions, workers iteratively implemented and revised function implementations, reflecting contributions from several participants. 
Participants submitted 175 \texttt{Review} microtasks. In 82 of these, they accepted the contribution by giving it a rating of 4 or more stars. 
One participant reported wanting more feedback:
\begin{quote}"The feedback offered was helpful. But sometimes I would get 4 stars with no feedback, which was not helpful at all. I think it should be mandatory to write some feedback just so I can know where to improve." - (P5)\end{quote}
\par

\subsubsection{Speed of onboarding and contributing}
During the two sessions, participants worked for a total of 31.5 hours. Participants spent 21 hours on microtasks that were submitted, including 39\% of their time on \texttt{Implement Function Behaviors} microtasks and 27\%  on \texttt{Review} microtasks. The remaining time was spent familiarizing themselves with the \textit{Crowd Microservices} environment, completing the post-task survey, and working on microtasks that were skipped rather than submitted.

After participants completed the tutorials and began work, participants spent additional time familiarizing themselves with the environment. On average, participants submitted their first microtask after 24 minutes. 

Participants skipped 17\% of all microtasks. Participants used the skip feature for several purposes. In some cases, participants skipped multiple microtasks to find an easy one with which to begin. Participants skipped \texttt{Implement Function Behaviors} tasks roughly ten times more often than \texttt{Review} tasks. At the beginning of the study, two microtasks were skipped automatically by the \textit{Crowd Microservices} due to 15 minutes time limit. All other skips were made intentionally by participants. \par

The median completion time was approximately 4 minutes for \texttt{Implement Function Behavior} and 3 minutes for \texttt{Review} microtasks (Table~\ref{table:microtasksCompletedSkipped}). These completion times are similar to the approximately 5 minute median completion times of prior microtask programming systems~\cite{latoza2018microtask}, despite requiring developers to test, implement, and debug their behavior.




\subsubsection{Feasibility of implementing microservices}
After the participants finished implementing the microservice, the project was deployed to a hosting service by \textit{Crowd Microservices}. To assess the feasibility of using a behavior-driven microtask workflow to implement microservices, we investigated the success of the crowd in building an implementation consistent with the described behavior in the \texttt{Client-Request}. We first constructed a unit test suite, generating a set of 34 unit tests (written and visible only to the experimenters, not participants), which is publicly available as part of our replication package\footnote{https://github.com/devuxd/crowd-microservices-output}. Overall, unit tests for 79\% (27) of the behaviors passed, and unit tests for 7 of the behaviors failed. To investigate the causes of the failing tests, we examined the microservice implementation created by the participants. We found that four of the failures were caused by a defect in one function involving a missing conditional, and the three remaining failures were either due to defects with behaviors not implemented correctly or not implemented. After we fixed these defects, all unit tests passed.



To further assess the implementation of the microservice built by the crowd, we used the final code written by participants to build a functioning \textit{ToDo} application. We first 
used the deploy feature in \textit{Crowd Microservices} to deploy and host the microservice. We then implemented a \textit{ToDo} application front-end as a React application, using the deployed microservice as the back-end. We found that, apart from the defects we described above, the \textit{ToDo} application worked correctly.

\begin{figure}
\includegraphics[width=\columnwidth,keepaspectratio, clip]{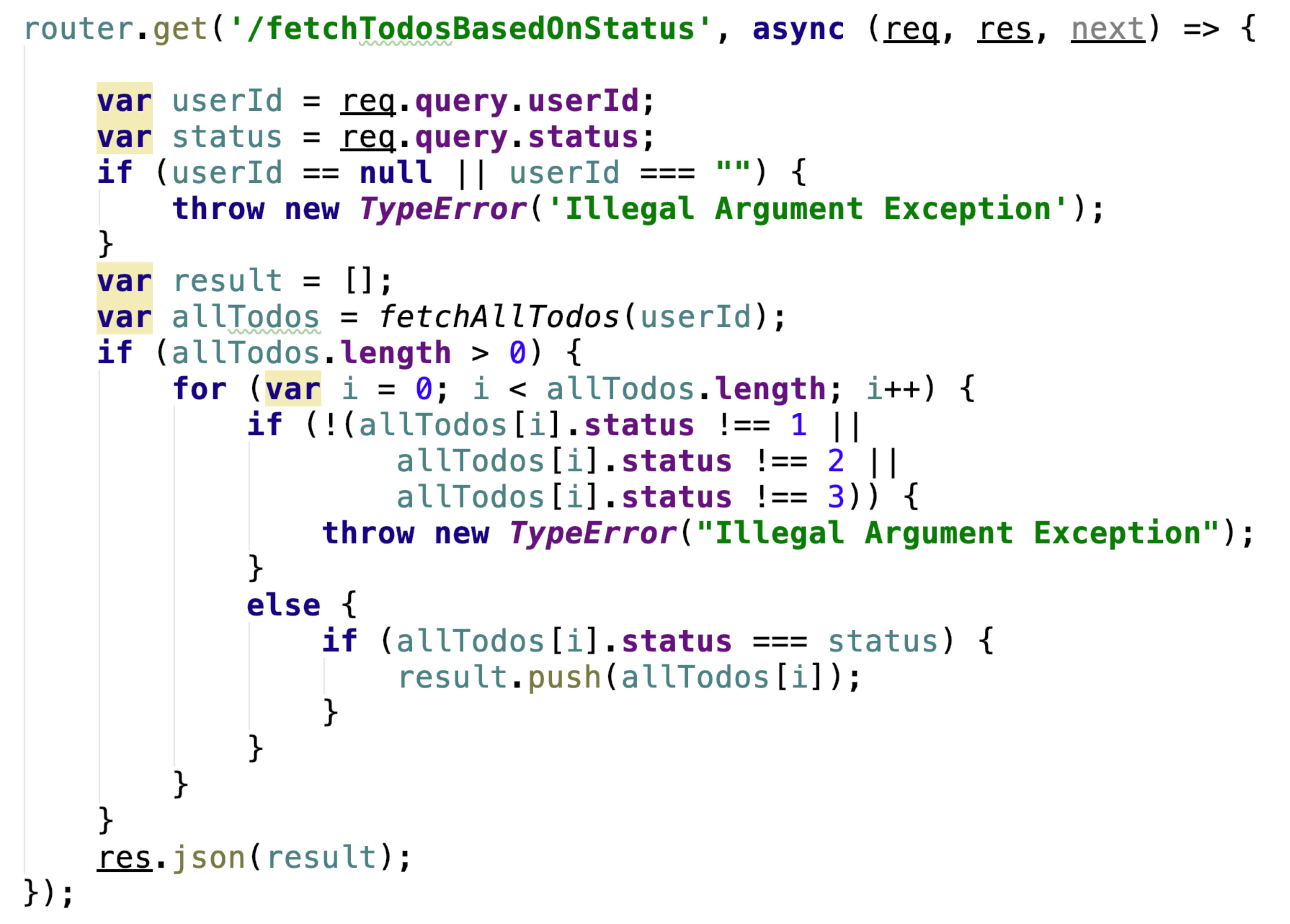}
\centering
\caption{An example of a microservice endpoint implemented by the crowd. 
}
  ~\label{fig:microSample}
\end{figure}

\textit{Crowd Microservices} offers an API for interacting with a persistence store, which participants made use of in their implementation. For example, in Fig.\ref{fig:microSample}, \texttt{fetchAllTodos} is an example of an invocation of the persistence API. Participants made a total of 15 calls to the persistence API, or 1.25 per function. 
In some cases, participants interacted with the persistence store indirectly, by calling other functions implemented by the crowd which made use of the persistence store. 
When asked in the post-task survey, most participants reported that they used the persistence API without any problems. Some participants reported that additional documentation would be beneficial: 
\begin{quote}"I used the API a little bit, and I felt like the documentation could be better with more examples." - (P5)\end{quote}

\section{Limitations and threats to validity}
Our study had several limitations and potential threats to the internal and external validity of the results. \par  

In our study, we chose to recruit a wide range of participants, recruiting participants locally from our university as well as globally through social networking sites. This yielded participants with a wide range of backgrounds, with their experience in JavaScript ranging from 2 months to 6 years. We chose this process as it mirrors the process of an open call, where contributors with a wide range of backgrounds may contribute. However, in practice crowdsourcing communities may exist in many forms, attracting many novice contributors looking for an entry point into more challenging work or attracting experienced workers who attract other experienced workers. Our results might differ if workers were exclusively more or less experienced. \par

Another potential threat to external validity is the choice of task. In selecting a task, we sought to identify a task that is representative of typical microservices that workers create. We chose the \textit{ToDo} application as a canonical example of a web application, often used to compare different approaches to building web applications. Larger microservices may involve more complex endpoints where individual behaviors are more challenging to identify. \par

Our results might also vary with different contexts in which work took places. To simulate the constant process of hand-offs that occur in microtask work, where workers complete tasks that others began, we chose to have participants work synchronously, maximizing the number of hand-offs that occur. To simulate participants coming back to work that they had begun earlier, we divided our study into two sessions. Of course, in practice, microtask work involves less predictable schedules, where contributors may come and go at arbitrary times. This may introduce additional challenges, where new participants that are unfamiliar with either the environment or anything about the project are constantly introduced. On the one hand, this might reduce performance, as such participants are less experienced. On the other, compared to participants in our study who had access to no workers who were already familiar with the environment and project, such participants might have an easier time onboarding, as more experienced workers would be available to answer their questions. Better understanding the impact of transient behavior on microtasking in programming is an important focus for future work. \par

In microtask work, workers are assumed to not have any prior information about the environment, the project, or other workers. However, in practice,  participants may over time gain experience with the project as well as the development environment itself. Over time, a contributor to our environment would experience fewer challenges using the environment and become more productive, reducing the average time for completing microtasks or increasing the amount of work contributors complete in each microtask.\par

 \section{Discussion}
Microtask programming offers a software development process in which large crowds of transient workers build software through short and self-contained microtasks, reducing barriers to onboarding and increasing participation in open source projects. 
In this paper, we explored a novel workflow for organizing microtask work and offered the first approach capable of microtasking the creation of web microservices. In our behavior-driven microtask workflow, each microtask involves a worker identifying, testing, implementing, and debugging an individual behavior within a single function. We found that, using this approach, workers were able to successfully submit 350 microtasks and implement 13 functions, quickly onboard and submit their first microtask in less than 24 minutes, contribute new behaviors in less than 5 minutes time, and together implement a functioning microservice back-end containing only 4 defects. Participants were able to receive feedback on their contributions as they worked by running their code against their tests and debugging their implementation to address issues.  \par


While our method has demonstrated success in implementing an individual microservice, there remain a number of additional challenges to address before a large software project could be built entirely through microtasks. The client is responsible for the high-level design tasks of determining the endpoints, designing data structures, and other design work. 
Moreover, while microtasking reduces the context a worker must learn to successfully contribute to a project, this context is not zero. Workers must still learn about the function they are working on and the current state of its implementation. This overhead is visible in the productivity data. In 21 hours working on submitted microtasks, 9 participants wrote only 216 lines of code and 397 lines of test code. \par


 Other work has explored techniques for decomposing design work into microtasks, such as through structuring work around tables of design dimensions and design alternatives\cite{Andre2016swDesign}. Such techniques might be adapted to a microtask programming workflow to enable the crowd to design the initial endpoints and data structures, as well as other high-level decisions, which might then be handed off to others who then implement this design. Similar to workflows such as TopCoder's, a senior crowd worker might also help ensure consistency across the design. Beyond upfront design, support is also necessary for maintenance situations where requirements change. This might result in changes in endpoints, data structures, and design decisions, requiring further downstream changes in the implementation. This might be addressed through new microtasks which identify changes, map these changes to specific impacted artifacts, and ask workers to update the corresponding implementation.\par

A key advantage of microtask programming approaches which incorporate automatic microtask generation is the potential to scale to large crowds and incorporate synchronous contributions from thousands or tens of thousands of developers. Rather than requiring a single worker acting as a client to manually generate each microtask, microtasks are generated automatically as the crowd works. Rather than potentially exposing contributors to the entire codebase and all of its ongoing changes, contributors must only understand an individual function. In this way, in principle, large crowds might be able to work together to build large applications quickly. For example, if a microservice ultimately resulted in 1,000 behaviors being identified, each behavior could then be worked on by a separate worker in a separate microtask. To the extent that these 1,000 microtasks can be done in parallel, this would then enable software development to occur with 1,000 concurrent microtasks, dramatically decreasing the time to complete work. Of course, many sequential dependencies might still exist, where, for example, the necessary existence of a function is not revealed until a previous function has already been implemented. Understanding just how many sequential dependencies exist in software development work and how much parallelism is truly possible is thus an important focus for future work.\par

While we developed our system in the context of implementing microservices, our approach could be adapted to apply to other types of software. For example, rather than beginning work with a set of endpoints, work might instead be begun with publicly visible methods in an API. Our technique could also be applied to building interactive applications, such as front-end web applications. However, such applications often make more extensive use of global state to, for example, render output or keep track of shared data entered by users. This raises additional challenges in effectively coordinating microtasks with only a local view of the system, requiring new techniques for surfacing appropriate information to workers.  \par 

There are many mechanisms for achieving quality in crowdsourcing systems. One common approach in Mechanical Turk crowdsourcing systems is replication, where multiple workers do the same work and a voting mechanism is used to aggregate these contributions~\cite{Doan:2011:www}. In microtask programming, this might used to review contributions. 
This approach has the advantage of being less susceptible to incorrect or poor quality reviews. But it requires additional work to generate these reviews. Another approach is to rely more on iteration. Even if a contribution is incorrect, subsequent contributions can update the artifact to address it. But there may be more effective ways to combine these approaches, such as offering contributors the ability to appeal reviews they perceive to be of low quality and ask for replicated reviews. \par

A wide range of volunteers, paid developers, or contract developers might participate in microtask programming projects. Our approach could be used to support open source projects. Studies have identified several motives for developers to join open-source projects, including a desire to learn and develop new skills, share knowledge and skills, and participate in a new form of cooperation\cite{ghosh2005understanding}. However,  developers face barriers to joining these projects, including  installing necessary tools, downloading code from a server, identifying and downloading dependencies, and configuring the build environment \cite{STEINMACHER2015,joiningOSS_Alex2003,Onion_Patch_Jergensen_2011,fagerholm2014onboarding}.
Consequently, onboarding  can require weeks,  discouraging casual contributors from joining. \textit{Crowd Microservices} may reduce these barriers by providing a preconfigured environment that helps developers onboard quickly. In our study, we found that developers could submit their first microtask in less than 24 minutes. 

Microtask programming might also be used by companies to develop software. While crowdsourcing involves recruiting contributors from outside a company or organization, it may also be possible to apply crowdsourcing inside a company. For companies with closed-source code and confidential information and intellectual property to protect, this model offers many of the potential crowdsourcing benefits of lower onboarding costs and greater resource fluidity with fewer of the potential drawbacks. \par

In adopting behavior-driven development, our approach benefits from the shared understanding among developers it can create. Creating this shared understanding is particularly critical in crowdsourced development, as there is no synchronous face-to-face communication to help synchronize understanding. 
As used in behavior development, this understanding may be created through the act of writing unit tests which makes shared knowledge concrete. Our approach also employs other techniques, such as a question and answer system. As the client through the client request has already made some of the key design choices, there may also be less need to coordinate, as there may be fewer design decisions that the crowd must make. Exploring techniques for achieving shared understanding of crowd workers is an important area for future work.  \par


\section{Conclusion}
In this paper, we offer a novel behavior-driven approach for microtasking programming work and apply this approach to implementing microservices. Our results offer initial evidence for its feasibility. Workers were able to successfully complete a microtask which asked them to identify, test, implement, and debug a behavior in less than 5 minutes as well as onboard onto a project in less than 30 minutes. Together, these contributions were able to be aggregated into a functioning microservice that was implemented entirely through microtask contributions. Important future work remains to investigate how this approach might be incorporated into a larger software project as well as exploring how a higher degree of parallelism might reduce the time to market in building software.\par

\bibliographystyle{cas-model2-names}

\bibliography{CrowdMicrorservicesref}

\bio{figs/Emad_headshot}{Emad Aghayi}
 received a BS degree in information technology from the Shiraz University of Technology in 2010 and an MS in information technology at the University of Tehran in 2014. He is currently a Ph.D. student in the Department of Computer Science at George Mason University. He is a member of the Developer Experience Design Lab, which studies how humans interact with code and designs new ways of building software. He works at the intersection of software engineering and human-computer interaction.  His research goal is to make developers more productive by providing better development tools and approaches. To achieve this goal, his research involves understanding developer behaviors, identifying problems, and designing tools and approaches to help with those problems using various HCI methods at each stage. Specifically, he researches and designs on the crowdsourcing in software engineering.
\endbio

\bio{figs/ThomasLaToza}{Thomas D. LaToza} received a Ph.D. in software engineering from Carnegie Mellon University in 2012 and degrees in psychology and computer science from the University of Illinois, Urbana-Champaign in 2004. He is currently an Assistant Professor of Computer Science at George Mason University. He serves as director of the Developer Experience Design Lab, which studies how humans interact with code and designs new ways of building software. He has conducted over 20 studies of software developers and designed numerous programming tools, including tools for designing, understanding, reusing, editing, and debugging code. His recent work has focused on programming environments which crowdsource insights. He served as co-chair of the Workshop on Crowdsourcing in Software Engineering and the Workshop on the the Evaluation of Programming Languages and Tools. In 2019, he was awarded an NSF CAREER award to support his research and teaching on debugging mental models. 
\endbio

\bio{figs/paurov}{Paurav Surendra}
 received a Masters degree in Software Engineering at George Mason University, Virginia and his Bachelors degree in Information Science and Engineering at Visvesvaraya Technological University in India. He is currently working as a Software Engineer at the Student Opportunity Center. His interests are in Software Architecture, Software Design, and Software Development for Web Platforms. 
\endbio
\bio{figs/meysam}{Seyedmeysam Abolghasemi} received a Masters degree in Computer Science at Old Dominion University in 2017 and a Bachelors degree in Information Technology and Systems from Monash University in 2012. He is currently working as a Senior Software Developer at Old Dominion University. His area of interests are Software Architecture and High-Performance Computing. 
\endbio

\end{document}